\documentclass[twocolumn]{aastex61}

\usepackage{amsmath}
\usepackage{gensymb}

\shorttitle{The Proper Motion of Segue 1}
\shortauthors{FRITZ et al.} 
\begin{document}

 \title{The Orbit and Origin of the Ultra-faint Dwarf Galaxy Segue~1}
 \correspondingauthor{T.~K.~Fritz}
\email{tfritz@iac.es}
\author[0000-0000-0000-0000]{T.~K.~Fritz} \affil{Department of Astronomy, University of Virginia, Charlottesville, 530 McCormick Road, VA 22904-4325, USA}\affil{Instituto de Astrofisica de Canarias, calle Via Lactea s/n, E-38205 La Laguna, Tenerife, Spain}\affil{Universidad de La Laguna, Dpto. Astrofisica, E-38206 La Laguna, Tenerife, Spain}
\author{M.~Lokken} \affil{Department of Astronomy, University of Virginia, Charlottesville, 530 McCormick Road, VA 22904-4325, USA}
\author{N.~Kallivayalil} \affil{Department of Astronomy, University of Virginia, Charlottesville, 530 McCormick Road, VA 22904-4325, USA}
\author{A.~Wetzel} \affil{Department of Physics, University of California, Davis, CA 95616, USA}
\author{S.~T.~Linden} \affil{Department of Astronomy, University of Virginia, Charlottesville, 530 McCormick Road, VA 22904-4325, USA}
\author{P.~Zivick} \affil{Department of Astronomy, University of Virginia, Charlottesville, 530 McCormick Road, VA 22904-4325, USA}
 \author{E.~J.~Tollerud} \affil{Space Telescope Science Institute, 3700 San Martin Drive, Baltimore, MD 21218, USA}

\begin{abstract}
We present the first proper motion measurement for an ultra-faint dwarf spheroidal galaxy, Segue 1, using SDSS and LBC data as the first and second epochs 
separated by a baseline of $\sim 10$ years. We obtain a motion of $\mu_{\alpha}\,\cos(\delta) = -0.37\pm0.57$ mas yr$^{-1}$ and 
$\mu_{\delta} =-3.39\pm0.58$ mas yr$^{-1}$. Combining this with the known line-of-sight velocity, this corresponds to a Galactocentric 
V$_\mathrm{rad}=84\pm9$ and V$_\mathrm{tan}=164^{+66}_{-55}$ km~s$^{-1}$. Applying Milky Way halo masses between 0.8 to 1.6$\times 10^{12}$ 
M$_\odot$ results in an apocenter at 33.9$^{+21.7}_{-7.4}$ kpc and pericenter at 15.4$^{+10.1}_{-9.0}$ kpc from the Galactic center, indicating 
Segue~1 is rather tightly bound to the Milky Way. Since neither the orbital pole of Segue~1 nor its distance to the Milky Way is similar to the more massive
classical dwarfs, it is very unlikely that Segue 1 was once a satellite of a massive known galaxy. Using cosmological zoom-in simulations of Milky Way-mass 
galaxies, we identify subhalos on similar orbits as Segue~1, which imply the following orbital properties: a median first infall 8.1$^{+3.6}_{-4.3}$ Gyrs ago,
a median of 4 pericentric passages since then and a pericenter of 22.8$^{+4.7}_{-4.8}$ kpc. This is slightly larger than the pericenter derived directly 
from Segue 1 and Milky Way parameters, because galaxies with a small pericenter are more likely to be destroyed. Of the surviving subhalo analogs only 
27\% were previously a satellite of a more massive dwarf galaxy (that is now destroyed), thus Segue 1 is more likely to have been accreted on its own.
\end{abstract}


\section{Introduction}
\label{sec:intro}

Until recently, 
athe nature of dwarf galaxies and globular clusters were clearly distinguishable by appearance alone. However, in the last decade, newly discovered objects have blurred the border between these two categories. These objects exist in a region of size-luminosity space previously uninhabited by globular clusters or dwarf galaxies \citep{Walsh_08}. One of the first discoveries of this kind was Segue~1, found by \citet{Belokurov_07b} using SDSS \citep{York_00}. It was first classified as a globular cluster, primarily because it was more compact and fainter (M$_V=-1.5$) than the galaxies known at this time. However, after extensive spectroscopic studies, this classification was disputed. \citet{Geha_09,Martinez_11,Simon_11} obtained low resolution spectroscopy of 394 stars in the field, determining that roughly 70 of those were probable Segue~1 members. \citet{Simon_11} measured a significant dispersion and derived an M/L$_V=3400$. They also measured a low metallicity of [Fe/H]$\approx-2.5$ with significant spread in the iron abundance. The spread in the [Fe/H] is robust and also confirmed by high resolution spectroscopy \citep{Frebel_14}. The consensus from these studies is that Segue~1 is a galaxy under the definition of \citet{Willman_12}, in which an iron spread is only possible in galaxies but not in globular clusters. If so, it is one of the closest galaxies to us, at a distance of $23\pm2$ kpc from the sun \citep{Belokurov_07b}.
However, the metallicity argument is not accepted by all; some like \citet{Dominguez_16} argue that Segue~1 could be a star cluster close to tidal disruption. 
In addition to its classification, another debated topic is whether or not the previously determined dark matter content of Segue~1 is correct or inflated. An accurate measurement of the dark matter content would be useful, as Segue~1 is probably the best dwarf galaxy for upper limits on dark matter decay due to its closeness and relatively high mass \citep{Scott_10,Aleksic_11,Aleksic_14}.

Since 2007, more galaxies similar to Segue 1
have been discovered (see for example \citet{McConnachie_12,Koposov_15,Bechtol_15}), but Segue~1 remains one of the faintest. In contrast to others like Cetus II \citep{Drlica_15}, it is already extensively studied, making it a desirable target for additional research. Due to its proximity, it is not only possible to study some of its stars in more detail than in most dwarf galaxies, but also to measure its proper motion with ground based telescopes as has been done for globular clusters  \citep[e.g.,][]{Dinescu_99a,Fritz_15}. Until now, proper motion measurements only exist for relatively massive dwarf galaxies down to the luminosity of $M_V\approx-8.8$ like Draco \citep{Pryor_15}. No ultra-faint dwarf spheroidal has a proper motion measurement thus far. A proper motion is essential for knowing the orbit of Segue~1, which enables comparison with other Milky Way satellite observations and simulations.

In this paper we present the first proper motion measurement for Segue~1. Section~\ref{sec:dataset} describes the data used for this study. We detail the methods used for the proper motion measurement in Section \ref{sec:deriv_pm}. 
In Section~\ref{sec:origin} we use this proper motion to constrain the orbit of Segue~1 and compare it with other satellites of the Milky Way, as well as simulations, to constrain the origin and history of Segue~1.
We conclude in Section~\ref{sec:summary}.

\section{Imaging Data Set} \label{sec:dataset}

In this section, we describe the imaging used to measure the proper motion of Segue~1.
As in \citet{Fritz_15}, we use SDSS data for the first epoch and the Large Binocular Camera on the Large Binocular Telescope for the second epoch. 

\subsection{SDSS data} \label{sec:sdss}

The first epoch consists of data from the SDSS catalog. We use DR12 \citep{Alam_15}. 
The most important properties we retrieve are the positions and their uncertainties. The uncertainties only include statistical uncertainty, so in Section~\ref{sec:distortion} we obtain the relevant systematic uncertainties from the data. We retrieve all objects from the PHOTObj tables in the SDSS CAS. 
Our selected R.A. and Dec. range covers the full area covered by LBC imaging with generous margins around it. Apart from positions, our query includes the following properties: {\tt PSF} magnitudes and their uncertainty for all bands, MJD, and {\tt probPSF} for all bands. The latter is used by SDSS to distinguish stars from galaxies. The MJD on which these observations were taken ranges from 53500 to 53766.

\subsection{LBC Imaging}\label{sec:lbt}

\begin{figure}
 \begin{center}
  \includegraphics[width=0.90 \columnwidth,angle=-90]{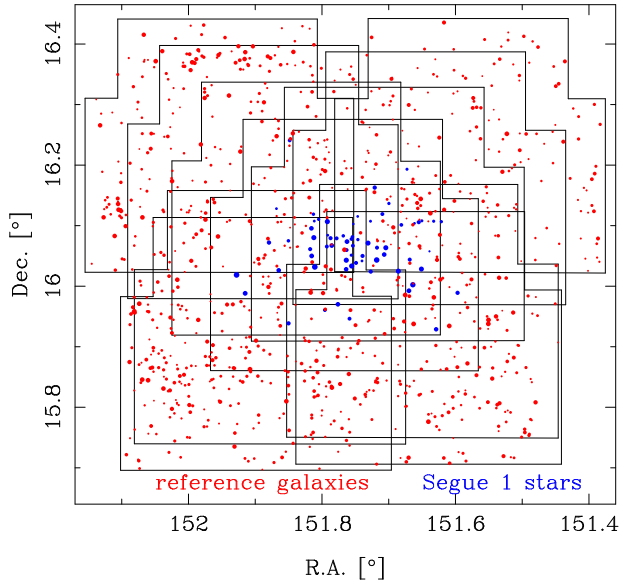} 
 \caption{Coverage of Segue 1. The black outlines show the coverage of the 11 used images. (The small chip gaps are not shown.) The blue dots show the 66 likely Segue~1 members and the red dots show the reference galaxies used for distortion correction and image registration. The area of the dots is antiproportional to the position uncertainties. 
 } 
 \label{fig:cov}
 \end{center}
 \end{figure}

The second epoch imaging is obtained with the Large Binocular Camera (LBC) \citep{Giallongo_08} at the Large Binocular Telescope (LBT). This camera covers a field of about 23' $\times$ 25' with four chips of 7.8'$\times$17.6', one situated above three others. For this work we use only the data best matching the SDSS data for astrometry. 
Since the primary band for SDSS astrometry is r-band, we use the r-filter (obtained with the blue eye of LBC) for LBC astrometry. We obtained 13 images, each exposed for 240 seconds, on the 8$^\mathrm{th}$ of February, 2016 (MJD of 57427). Two images are badly focused, as addressed in Section~\ref{sec:distortion}, and thus are not used in the analysis.
The images were dithered to produce the pattern shown in Figure~\ref{fig:cov}. The data set gives a baseline of about 10 years relative to SDSS. The pixel scale is 0.226$\arcsec$ and the typical FWHM is 0.94$\arcsec$. Thus the PSF is more than Nyquist sampled and most galaxies are resolved. We reduce the data in the usual way using custom scripts\footnote{For the reduction we use dpuser as for many basic calculations: \url{http://www.mpe.mpg.de/~ott/dpuser/index.html}}, beginning by constructing a flat with sky images. We then apply it to the bias-corrected sky frame and interpolate over permanently bad pixels. This reduction is sufficient for our purposes because we are not interested in extended low surface brightness features. Cosmic rays are dealt with later when they affect a relevant object in one image; since we have several images (up to 10) of the important area, we can recognize and reject outliers. We do not combine the different images into one stack, because when a stack is created without a good knowledge of distortion, it is very difficult to correct for residual distortion later \citep{Gillessen_09}.

\section{Proper motion measurement} \label{sec:deriv_pm}

This Section describes the process of identifying object properties from image data, classifying sources, and correcting positions for differential chromatic refraction and distortion. These procedures lead to the proper motion measurement and Galactocentric velocities in Sections \ref{sec:proper_motion} and \ref{sec:motionfinal}. Most methods in this Section are similar to the approach in \citet{Fritz_15}.

\subsection{Measuring pixel position in LBC images} \label{lbcpix}

We use SExtractor \citep{Bertin_96} to find and measure objects on the LBC images. Each chip is treated separately. We use mostly standard parameters,
but the saturation limit is changed to 60000 ADU and the seeing FWHM to 0.9$\arcsec$. The sources which are detected in the SDSS are also very well detected in the deeper
LBC images when the standard threshold of  5 pixels above 1.5 $\sigma$ is used.
 SExtractor measures several properties; the most important for our work are the various techniques to measure the position. For position, we choose to use the measurements {\tt XWIN}/{\tt YWIN} since they are relatively assumption-free and are much more precise than {\tt X}/{\tt Y} \citep{Fritz_16}. 
For the position uncertainty we use {\tt ERRX2WIN\_IMAGE}/{\tt ERRY2WIN\_IMAGE}. 
We also run PSFex \citep{Bertin_11} mainly to obtain information about the morphology of our sources. We mainly use the default configuration settings,
 with the exception of a 
{\tt SAMPLE\_FWHMRANGE} of  2.0  to 10.7 pixels and a  
{\tt SAMPLE\_VARIABILITY} of 0.2. This wide range allows PSFeX to consider all sources as initial candidates for building the PSF model before later removing sources
based on S/N and goodness-of-fit.
These PSFs are then used in the second SExtractor run to obtain {\tt FLUX\_PSF}  and {\tt FLUX\_MODEL},  their errors,  and {\tt SPREAD\_MODEL}. The latter is used to distinguish galaxies from stars.

\subsection{Initial Transformation, Source Matching, and Magnitude Calibration} \label{sec:in_trafos}

To make source-matching possible between the two data sets, position information for one must be transferred approximately to the same frame as the other. The source positions from SDSS are already in sky coordinates and astrometrically calibrated by \citet{Pier_03}. LBC source positions, currently in pixel space, must undergo coordinate transformations and distortion corrections before being comparable to SDSS. The distortion solution and image registration from \citet{Fritz_15}, derived for LBC r-band data from the LBT red eye, provide a starting transformation. Next, we perform iterative additional transformations using well-defined LBC sources as references, matched to their SDSS counterparts by position and PSF magnitude criteria. Initially only sources near each chip center are used; subsequent transformations include sources further and further out. This culminates in a cubic transformation which includes sources from the full chip.

The sources found in the LBC data are then matched with nearest neighbor sources in SDSS through similarity in position and PSF magnitude. However, more precise object comparison necessitates a magnitude calibration between the data sets. We scale LBC magnitudes per chip, for each image, to the SDSS system by comparing stars brighter than the SDSS r-band faint limit with their LBC nearest neighbors. Each of the stars used is classified as a star by SDSS's $\tt prob\_PSF$ property in all three g, r, and i bands. For each star, we find the difference between $-2.5 \log({\tt FLUX\_MODEL})$ and the SDSS $\tt ModelMag$. After a three-sigma cut, the average of this difference becomes that image's zero-point for LBT magnitudes.

\subsection{Source Classification} \label{sec:sources}

Since our proper motion is measured relative to background galaxies, we need a clean sample of these for a stable reference frame. Therefore,  we apply the following criteria for galaxy selection. First, the closest matched SDSS source needs to be classified as a galaxy in all three high SNR bands (g, r, and i). This means the SDSS classifier $\tt prob\_PSF$ is 0 for all three. In addition, the object needs to be a galaxy on the LBC image; we use ${\tt SPREAD\_MODEL}$ from SExtractor and require $|{\tt SPREAD\_MODEL}|>0.003$ \citep{Desai_12}.
All other objects are treated as stars. This definition of stars is less clean, but a clean sample of stars is less important, because we use spectroscopic information to define the Segue~1 sample in Section~\ref{sec:selection}. This object classification is also used for the correction of differential chromatic refraction in Section~\ref{sec:dcr}.

\subsection{Segue~1 star selection} \label{sec:selection}

\begin{figure}
 \begin{center}
   \includegraphics[width=0.70 \columnwidth,angle=-90]{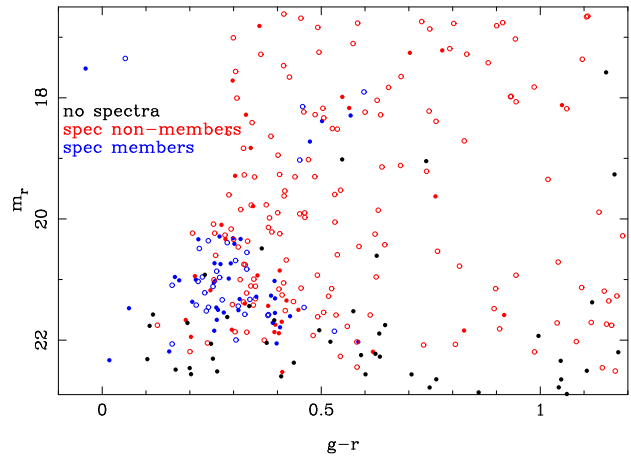} 
 \caption{Color magnitude diagram of stars in Segue~1. Shown are all SDSS stars (classified as stars by SDSS in at least two of the gri bands)  within the half-light radius (4.4') as solid dots. It is visible that nearly  all relatively bright, blue stars have spectroscopy in the \citet{Simon_11} sample. The majority of these within the half-light radius are identified as spectroscopic members in their subjective classification (blue), while some are not members (red). In addition, we show the stars at larger distances with spectroscopy as open circles. The color and magnitude range is reduced to about the range of the Segue~1 members.
 } 
 \label{fig:members}
 \end{center}
 \end{figure}

Our observations contain both Segue~1 stars and unassociated field stars. Usually member stars are selected mainly in color magnitude space--see e.g. \citet{Fritz_15,Dinescu_16} 
with spatial \citep{Fritz_15} and/or astrometric \citep{Fritz_17} information sometimes used in addition.
For satellites like Segue~1 which are not very prominent in surface density, these criteria are not ideal because they do not result in clean samples when the proper motions of single stars are not very precise. For Segue~1 we have the possibility to get a cleaner sample by employing the spectroscopy of \citet{Simon_11}. This sample should be 98.2\% complete down to an r-magnitude of 21.7 within a radius of 10' (2 half-light radii, which covers nearly our full field of view), see also Figure~\ref{fig:members}.  
 
 Of the 390 stars of \citet{Simon_11} we find matches for 311. The missing stars are mostly outside the field of our coverage. \citet{Simon_11} give three ways to classify stars as members. The first is a subjective criterion, which is binary. Of the 71 stars classified as members ($p=1$) by the binary method, only 2 are not in our initial sample. However, we exclude stars brighter than 18 magnitude in r-band from the sample, since bright stars are saturated in LBC (see Section~\ref{sec:systs-tests}). Excluding these stars results in a sample of 66 stars.
 Two other criteria (expectation maximization, or EM, and a Bayesian approach), give a probability for each star being a Segue 1 member\footnote{Photometric non-members have a probability value of -9.999 in \citet{Simon_11}. We treat them as if they have a value of 0 in the following.}. Thus, even with spectra the membership is not certain. With three different member definitions we can estimate the impact of membership uncertainty.
In the case of the subjective criterion, obtaining the sample is trivial: we simply use all stars with $p=1$. In the other two cases we run Monte Carlo simulations to get samples: each \citet{Simon_11} star is given a random number between 0 and 1. When the random number is smaller than the probability of the star, the star is added to the sample. With that algorithm a star with a membership probability of x, is with a likelihood of x in the sample.  We repeat this process 100 times, which results in 100 different samples for each of the two cases. On average, we use 68 stars in the EM case (the range goes from 65 to 71) and 64 in the Bayesian case (the range goes from 58 to 66 stars). We use all of the 201 (100 EM cases, 100 Bayesian cases, 1 subjective) different samples in Section~\ref{sec:distortion} to get a proper motion for each case.

\subsection{Differential Chromatic Refraction} \label{sec:dcr}
The Earth's atmosphere refracts incoming light towards the normal, causing objects to appear closer to the zenith than their true location. The refraction angle $\alpha$ is given by

\begin{equation}
\alpha = \alpha^{'}\tan{\zeta},
\end{equation}

where $\zeta$ is the source's zenith angle and $\alpha^{'}$ is the deflection at $\zeta=45^{\circ}$. Refraction is also dependent on the source's wavelength range. Differential Chromatic Refraction (DCR) causes blue light to appear closer to zenith than red, presenting a problem for relative astrometry. This effect is contained in $\alpha^{'}$:

\begin{equation}
\alpha^{'} = \frac{n(\lambda)^2-1}{2n(\lambda)^2}.
\end{equation}

Here $\lambda$ is the object's effective wavelength in the band used for observation, and $n$ is the index of refraction. 

While most of the refraction is corrected by linear transformations in Section \ref{sec:distortion}, DCR offsets must be separately accounted for \citep{Fritz_09}. The SDSS data is already partially corrected for DCR as part of the \citet{Pier_03} astrometric calibrations; however, their linear relation between object color and relative altitude shift does not fit objects as precisely as needed here. We remove the SDSS correction and instead apply the equations derived in Section~3.2 of \citet{Fritz_15}, which relate $r-i$ color with relative shift to zenith. These relations differ for stars versus galaxies, so we apply the appropriate corrections separately using the classifications for stars and galaxies in Section~\ref{sec:sources}. 
In the application of the DCR correction, the source positions are converted from equatorial to horizontal coordinates and the DCR relative shift in altitude is subtracted from all SDSS and LBT object positions.

\subsection{Distortion correction, image registration and position uncertainties} \label{sec:distortion}

A proper motion measurement needs a stationary reference frame against which positions from the two epochs are compared. SDSS data has already been astrometrically calibrated by \citet{Pier_03}.  To profit from that we use the same approach as \citet{Fritz_15}, in which the SDSS positions are taken as a distortion-free reference frame and LBC positions are shifted to that frame via a distortion correction. As in \citet{Fritz_15}, member stars and background galaxy positions are fit together, and the fit also determines the motion offset of the target. The motion offset is then converted into the proper motion by dividing by the baseline. 
In practice we start by only using galaxies in the fit so that we already have a selection of good background galaxies when we add the Segue~1 members, whose selection is more uncertain.

For the distortion correction, we select from the galaxies defined in Section~\ref{sec:sources} only those with m$_r<$21.5. Fainter galaxies have too low SNR in SDSS. In addition, we exclude galaxies with large position uncertainties as well as those which are more than 3$\arcsec$  distant from the closest SDSS match after the initial transformation. The latter are probably poor matches. 
Finally, for each image, we exclude galaxies from the sample when they are near the edge of a chip. Together, these requirements yield on average approximately 50 galaxies per chip per image. We assess the galaxy coverage of each chip by eye, verifying that there are no severe gaps where the transformation would be poor.

The following equations relate the corrected positions ($x_{cor}$ and $y_{cor}$)  to the original $x$ and $y$ pixel coordinates by the addition of a multivariate cubic polynomial. $x^{'}$ and $y^{'}$ are positions relative to each approximate chip center at (1049, 2304).

\begin{equation}
\begin{split}
\label{eq:distortion}
x_{cor} = x + a_1 x^{'2} + a_2 x^{'}y^{'} + a_3 y^{'2}  + a_4 x^{'3} \\
+ a_5 x^{'2} y^{'} + a_6 x^{'} y^{'2} + a_7 y^{'3} \\
y_{cor} = y + b_1 x^{'2} + b_2 x^{'} y^{'} + b_3 y^{'2} + b_ 4 x^{'3}\\
+ b_5 x^{'2} y^{'} + b_6 x^{'} y^{'2} + b_7 y^{'3}
\end{split}
\end{equation}
 \citet{Fritz_15} found that using higher-order polynomials does not improve the residuals. The same coefficients are fit to all images, since distortion is constant on the time scale of a night or even longer \citep{Fritz_09,Fritz_15}. Meanwhile, linear terms are encompassed by the transformation to sky coordinates:
 
\begin{equation}
\begin{split}
\label{eq:linear}
R.A. = c_1 + c_2 x_{cor} + c_3 y_{cor}+c_{4segue}\\
Dec. = d_1 + d_2 x_{cor} + d_3 y_{cor}+d_{4segue}.
\end{split}
\end{equation}

Most parameters of Equation~\ref{eq:linear} are fit again for each image since they change mainly due to dithering and varying airmass. The parameters $c_{4segue}$ and $d_{4segue}$ are fit to all images together. They are the offset of the Segue~1 members relative to the reference galaxies due to the proper motion of Segue~1. The four chips are treated completely separately in that approach.
We also test variants of the equations, which involve coupling of the four chips. In these versions, inter-chip distances and relative orientations are additional parameters. In the first variant we fit each image separately and compare the motion obtained from the different images. The first two images give significantly different motions than the others, and upon further inspection are found to have clearly worse focus than the others; therefore we do not use these images.
 As another version, we also test fitting all images and four chips at once. The inter-chip distance and orientation here is the same for all images. This variant has a reduced $\chi^2$ enlarged by about 0.2 compared to the standard case, similar to our observation in \citet{Fritz_15}. Probably, the chip orientation and/or distance is not fully constant.

The optimal coefficients for Equations \ref{eq:distortion} and \ref{eq:linear} are fitted simultaneously for each of the chips using the ${\tt mpfit}$ package \citep{Markwardt_09}.
The position uncertainties ($\sigma_{ra/dec}$) input to the fit are the SDSS-given $\tt raErr$ and $\tt decErr$ enlarged by unknown values $add$ and $fact$:

\begin{equation}
\label{eq:error}
\sigma_{ra/dec} = \sqrt{(ra/decErr \times fact)^2 + add^2}.
\end{equation}

The value $add$ will fold in the systematic SDSS uncertainties which are not included in the given position uncertainties, and $fact$ can account for LBT uncertainties or SDSS random uncertainty underestimates. For the initial run, we use $add=17.21$ mas and $fact=1.186$, the results from \citet{Fritz_15}.

In the initial runs we fit only galaxy positions but not Segue~1 members. After the first run, we remove galaxies with high residuals from the sample of references. We first remove outliers with $R>7$, where $R=\sqrt{R_x^2+R_y^2}$ and $R_{x,y}$ are the differences between a galaxy's transformed LBC position and SDSS position divided by the uncertainty from Equation~\ref{eq:error}. The eliminated sources probably consist of those which were misclassified, contained bad pixels or had complex morphologies causing centering uncertainty. After a second fit with the reduced sample, we repeat the cut with $R>$5. The initial higher residual cut is meant to include galaxies which were fine for use as references but poorly affected by the worst sources in the first run.

We then apply the derived distortion solution to all objects. We check the distortion corrected positions of other objects, of which the Segue~1 members are most important, for outliers. 
To identify outliers, we calculate for all objects with more than 2 detections the standard deviation of position for all detections and compare it with the standard deviation when one detection is omitted. When the standard deviation is more than a factor of 2 
smaller with one detection omitted, we classify that as an outlier and omit it from the average. The factor of 2 results in outlier rejection for 11\% of the objects. For objects with less than 6 detections, the factor of 2 is increased to a larger factor to produce the same fraction of outliers, since in small samples factors of the order 2 can happen by chance. We also check different factors; however, as they change the final motion by less than 0.04 mas/yr, which is about 1/15 of the uncertainty, outliers are not important and we do not test them further.

\begin{figure}
 \begin{center}
     \includegraphics[width=0.70 \columnwidth,angle=-90]{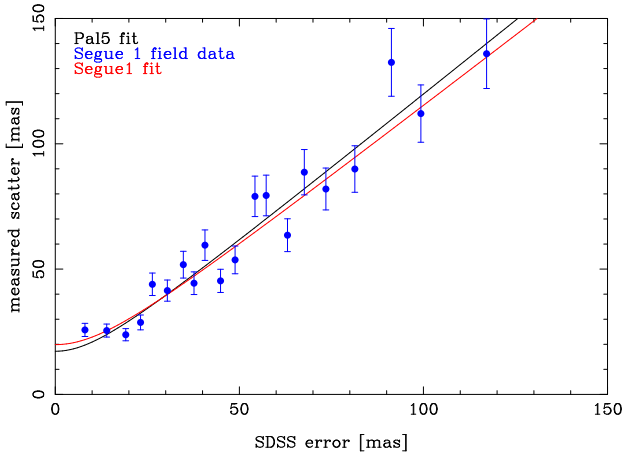} 
 \caption{
 Errors of the reference galaxies. The y-axis shows the robust scatter of their offsets, the difference in SDSS and LBC positions. The x-axis shows the binned SDSS uncertainties. The fits plotted show the uncertainty term used for the transformation of a source for a given SDSS uncertainty. It is visible that the fit is very similar to the fit for the Palomar 5 field of \citet{Fritz_15}.
 } 
 \label{fig:errors}
 \end{center}
 \end{figure}

To determine the $add$ and $fact$ (Equation~\ref{eq:error}) for this sample, we use the method from \citet{Fritz_15}. Galaxies are binned by their SDSS-given R.A./Dec. uncertainty into twenty equally-populated groups, and the scatter of the offsets (for which we use a robust measure of $1.483$ $\times$ the median deviation) is calculated for each, where the offset is the difference in position between the SDSS position for the galaxy and the LBC position in each individual image.
We fit the resulting data with the initial guesses for Equation~\ref{eq:error} and find a fit very similar to the Palomar~5 parameters, see Figure~\ref{fig:errors}. The new parameters are $add=19.85$ mas and $fact=1.136$. The new parameters reduce the $\chi^2$ by 2.5. We perform a new run of the distortion correction with these updated uncertainties. The influence on the motion is minimal; thus no further iterations are necessary. While these uncertainties are correct for a single object, they are not optimal for our purposes because many objects (especially the Segue~1 stars close to the center of the field) are detected several times by LBC. However, these multiple detections are matched to a single SDSS source and are always associated with the same SDSS position and uncertainty. Using only the aforementioned uncertainty would give them too much weight. Instead we increase the uncertainty of all sources used in the fit by $\sqrt{N_\mathrm{detect}}$, where $N_\mathrm{detect}$ is the number of detections of each source counting over all four chips. The change in the motion by the modified uncertainties is clearly smaller than the final uncertainty. 
Thus, the impact of the choice of uncertainties on the final motion is minimal.

\subsection{Tests for systematics}\label{sec:systs-tests}

Here we test for systematics by plotting the motions of the Segue~1 member stars against different properties. For this we use the subjective member classification of \citet{Simon_11}, since these stars were determined to be members with high certainty and using less certain members might make trends invisible. The distortion solution of Section~\ref{sec:distortion} produces a motion for each source on each LBC image, which we average over all images to get a single motion for that source. The position uncertainty of each object comes from Equation~\ref{eq:error}. However, we enlarge the uncertainty of each object by the same factor which is chosen such that the uncertainty of the average motion of Segue~1, obtained by error weighted averaging of all members, is the same as obtained in our final proper motion in Section~\ref{sec:proper_motion}. This is necessary because just using the uncertainties of Segue~1 members ignores other sources of uncertainties, like position uncertainties of galaxies, and underestimates the total uncertainty.

\begin{figure*}
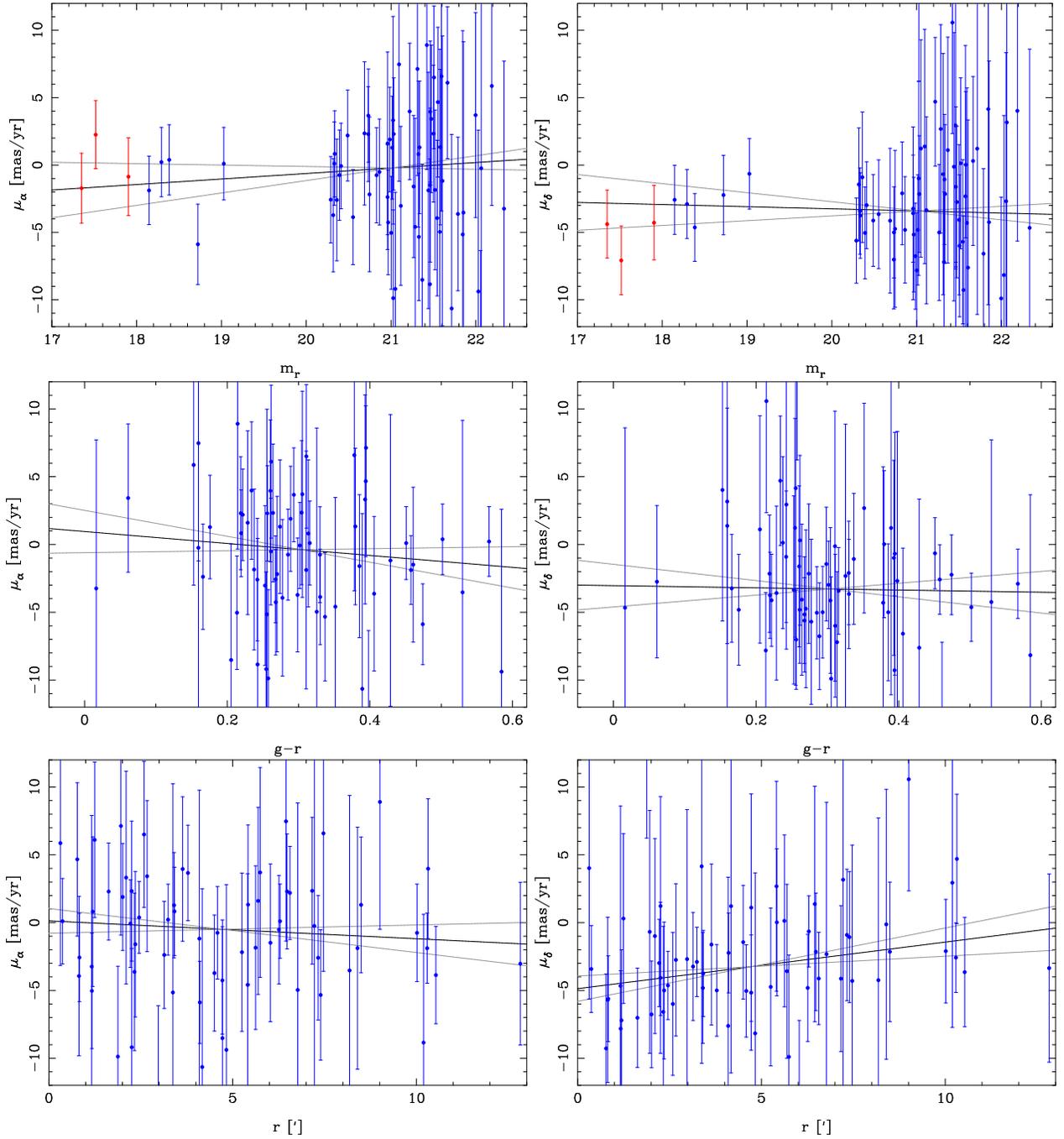

 \begin{center}
    \includegraphics[width=0.70 \columnwidth,angle=-90]{f4a.eps}
     \includegraphics[width=0.70 \columnwidth,angle=-90]{f4b.eps}
    \includegraphics[width=0.70 \columnwidth,angle=-90]{f4c.eps}
     \includegraphics[width=0.70 \columnwidth,angle=-90]{f4d.eps}    
         \includegraphics[width=0.70 \columnwidth,angle=-90]{f4e.eps}
     \includegraphics[width=0.70 \columnwidth,angle=-90]{f4f.eps}       
 \caption{Fits to the proper motion as a function of r-magnitude, $(g-r)$ color, and distance of the cluster center. All blue points indicate Segue 1 members used in the final proper motion calculation and the fits. The red points stand for stars brighter than m$_r=18$ that are not used in the final calculation or the fits. Each point on the plots represents the given source's proper motion and respective quantity. The fit linear functions, shown in black with the 1 sigma width marked in gray, do not reflect the final calculation of the proper motion but rather are to display any potential dependencies in the motions. Within the uncertainties, there do not appear to be significant trends in the proper motions of the Segue 1 stars.  
 } 
 \label{fig:syst}
 \end{center}
 \end{figure*}

We test for trends of $\mu_\alpha$ and $\mu_\delta$ as function of R.A., Dec., the distance from the uncertainty-weighted center of the Segue~1 member stars (151.76,16.05), m$_r$, and $(g-r)$. 
None of these trends are larger than 1.74 $\sigma$, and the average is 0.66 $\sigma$. We show the proper motion as function of $(g-r)$, m$_r$  and radius in Figure~\ref{fig:syst}. Stars brighter than 18$^{th}$ magnitude are offset in $\mu_\delta$ because because they contain several saturated pixels in the LBC images.
 Therefore we do not use stars brighter than m$_r=18$ for this work. In the case of slightly fainter stars, one or two pixels are saturated only in a subset of images, too few to affect the centering (as also visible in the Figure). Thus, we still include these stars. Additionally, the Figure shows no trend as a function of $(g-r)$. That indicates that our DCR correction works. The largest trend of 1.74 $\sigma$ is the trend of $\mu_\delta$ as a function of the radius, but one trend of that size is expected when 10 different trends are checked, since in a Gaussian distribution a value of 1.74 $\sigma$ or larger happens with 8.2\% probability. 

\subsection{Proper Motion and its Error} \label{sec:proper_motion}

Our fits in Section~\ref{sec:distortion} directly obtain the offset between Segue~1 stars and the reference frame. We convert these offsets into proper motions by dividing by the average time baseline of 
10.4 years and by changing the sign.
The latter is necessary because we transform from the second epoch to the first, so the initial motion we measure is the inverse motion.
We also correct for $\cos{\delta}$ in the case of $\mu_\alpha$. All mentions of $\mu_\alpha$ in the following are in reality
 $\mu_\alpha\,\cos{\delta}$, but we omit $\cos{\delta}$ in the following. Thus we obtain the proper motion for our three cases (subjective, EM, and Bayesian, see Section~\ref{sec:selection} and \citet{Simon_11}) of Segue~1 member selection. In the case of the subjective probability, it is just the value of the single fit, while in the EM and Bayesian cases we average the motions obtained by the 100 Monte Carlo simulations to one value. The final value is then the average of the three cases.

The uncertainty of the Segue~1 proper motion due to member star selection is 
\begin{equation}\sigma_\mathrm{sel}=\sqrt{[\sigma(Avg_\mathrm{Methods})]^2+[(\sigma(N_\mathrm{EM})+\sigma(N_\mathrm{Bay}))/3]^2}.\end{equation} 
The error has two contributions, firstly the scatter over the different methods ($\sigma(Avg_\mathrm{Methods})$) and secondly the scatter within each method ($\sigma(N_\mathrm{EM})$  and $\sigma(N_\mathrm{Bay})$).  The second term is divided by 3, because there are three methods of star selection. One (the subjective) has no uncertainty, thus it does not contribute to the numerator of the second term.
It contributes 0.170/0.131 mas yr$^{-1}$ in R.A./Dec.
The uncertainty caused by the position uncertainties of Segue~1 stars and reference galaxies cannot be directly obtained by the fit, because many objects are used several times in the process. Instead we use the half-sample method \citep{Feigelson_12}. That means we obtain the uncertainty by comparing the motions of two independent subsamples. We create  two subsamples by randomly adding objects (member stars and reference galaxies) to one of the two. Since this is done on the object level, the object is omitted from all images in one subsample and included in all its images in the other. We then obtain the preliminary proper motion uncertainty of the two subsamples by averaging the two. Next we calculate the reduced $\chi^2$ of that case and then change the errors such that the  reduced $\chi^2$ has the expected value of 1. With the rescaled errors we again calculate the weighted average, which we use in the following.
We repeat this process of generating two subsamples 100 times with different random division into two subsets. The final uncertainty is the average of the proper motion errors of these 100 cases, which is 0.543/0.565 mas yr$^{-1}$ in R.A./Dec.
 We now test whether the motions in R.A. and Dec. are uncorrelated. For that we do not use the correlation matrix obtained by the fit because it underestimates the correlated errors. Instead we use again the 200 subsamples, which means we use the two subsets of the 100 realizations independently. Using {\tt fitexy.pro} with iterative outlier rejection, we fit a line to the $\delta$ motion as a function of the $\alpha$ motion and the errors of both values to obtain the slope, which is $-0.89$. 

The total uncertainty is the quadrature sum of these individual components, which is 0.57/0.58 mas yr$^{-1}$ in R.A./Dec. Since the other uncertainties are not correlated in the two dimensions, the overall slope is reduced to $-0.86$.
Altogether, we obtain a proper motion of  $\mu_\alpha=-0.37\pm0.57$ mas yr$^{-1}$ and $\mu_{\delta}=-3.39\pm0.58$ mas yr$^{-1}$.  In the conversion to Galactocentric velocity and other properties we also consider the correlation between the two dimensions.

\subsection{Galactocentric Velocity} \label{sec:motionfinal}

To convert the proper motion to physical units we additionally require both the distance to Segue~1 and the Solar position and velocity relative to the Galactic Center.
For Segue~1 we use a distance of $23\pm2$ kpc from the sun \citep{Belokurov_07b}. 
 
From the apparent motion of Sgr~A* \citep{Reid_04}, it is possible to infer the solar motion in the direction of Galactic rotation (V') when the distance to the Galactic Center is known ($R_{GC}$). 
The circular velocity at the position of the sun is not necessary for the conversion. 
 We adopt $R_{GC}$=8.2 kpc as determined in the review of \citet{Bland_16}. This is also consistent with most recent precise determinations like \citet{Dekany_13,Chatzopoulos_14,Reid_14,Gillessen_17}. 
 The solar velocity relative to the local standard of rest is well-determined in the radial (U) and vertical (W) direction \citep{Reid_04,Schoenrich_10,Bovy_12b}, and thus we use these directly.
 The resulting solar motion with respect to the Galactic center is U/V'/W$=$11.0/248.0/7.3 km~s$^{-1}$. 
 To obtain the full Galactocentric velocity of Segue~1, the heliocentric line-of-sight velocity also needs to be included, for which we use $208.50\pm0.9$~km~s$^{-1}$\citep{Simon_11}.

The uncertainties in the parameters for the sun are small compared to the uncertainties for Segue~1. We ignore the former in the following.
 To estimate uncertainties including correlations between them in Galactocentric positions and velocities\footnote{We use the same coordinate-system conventions as \citet{Kallivayalil_13}.} we add to the properties with the largest uncertainties (proper motions, distance, line-of-sight velocity) Gaussian random numbers with the width of the uncertainty. We draw random numbers 100,000 times and then calculate the difference between the median and the borders of the 1-$\sigma$ range for each parameter. This difference is used as error which can be asymmetric. 
 For the value we use the most likely proper motion, since the median of the Monte Carlo simulation can be biased, especially for positive definite properties.
 We obtain for the position: X/Y/Z: $-19.4\pm1.0$/$-9.5\pm0.8$/$17.7\pm1.6$ kpc and for the velocities V$_\mathrm{X}$/V$_\mathrm{Y}$/V$_\mathrm{Z}$:  $13\pm33$/$-175^{+70}_{-75}$/$51\pm51$
 km~s$^{-1}$. 
 The velocity is equivalent to Galactocentric V$_\mathrm{rad}=84\pm9$ km~s$^{-1}$ and V$_\mathrm{tan}=164^{+66}_{-44}$ km~s$^{-1}$
 \footnote{As usual in that context we use the definition in which tangential velocity cannot be negative.}. These and other properties are summarized in Table~\ref{tab:segsummary}.

 \begin{table*}
 \centering
 \caption{Summary of Segue~1 Properties. \label{tab:segsummary}}   
 \begin{tabular}{ l l l}
 \hline \hline 
 Property & Measurement  & Source \\
 \hline 
  R.A./Dec. 			& 151.766$^\circ$/16.08 $^\circ$						& \citet{Belokurov_07b}\\
  l/b 				& 220.5$^\circ$/50.4$^\circ$ 				& \citet{Belokurov_07b}\\ 
  Distance 			& $23\pm2$ kpc						&  \citet{Belokurov_07b} \\
  X/Y/Z 				& $-19.4\pm1.0$/$-9.5\pm0.8$/$17.7\pm1.6$ 	& this work\\
  $v_{los}$ 			&  $208.5\pm0.9$ km~s$^{-1}$ 							& \citet{Simon_11} \\ 
  proper motion 		& $-0.37\pm0.57$/$-3.39\pm0.58$ mas yr$^{-1}$ & this work\\
  V$_\mathrm{X}$/V$_\mathrm{Y}$/V$_\mathrm{Z}$ & $13\pm33$/$-175^{+70}_{-75}$/$51\pm51$ km~s$^{-1}$ & this work \\ 
  V$_\mathrm{rad}$/V$_\mathrm{tan}$/V$_\mathrm{tot}$ & $84\pm9$/$164^{+66}_{-44}$/$184^{+63}_{-38}$ km~s$^{-1}$  & this work \\ 
 \hline \hline
 \end{tabular}
 \end{table*}

\section{Orbital history and origin of Segue 1} \label{sec:origin}
  
In this Section we first obtain the orbit of Segue~1 and then compare it with the orbits of other satellites and predictions for the proper motion of Segue~1. Then we use simulations and other properties of Segue~1 to draw conclusions about its origin. 

\subsection{Orbit of Segue~1} \label{sec:orbit}

The orbit of Segue~1 depends not only on its current position and velocity but also on the potential of the Milky Way.
For the potential we use ${\tt MW2014}$ of \citet{Bovy_14b} 
using the software ${\tt galpy}$\footnote{\url{http://github.com/jobovy/galpy}} from the same work, which assumes a solar distance to the Galactic Center of $R_{0} = 8$ kpc. We verified with simple potentials that the influence of the differences in R$_0$ are small compared to the uncertainties of the potential.  
To explore the influence of the measurement uncertainties we generate 1000 subsamples, by adding Gaussian uncertainties of the width of the uncertainties to the proper motions, line-of-sight velocity and distance modulus. We obtain for the pericenter a distance of 17.9$^{+8.2}_{-9.6}$ kpc and for the apocenter 37.0$^{+29.7}_{-6.8}$ kpc. 
That corresponds to an orbital period of about 700 Myrs. 

\begin{figure}
 \begin{center}
   \includegraphics[width=1.08 \columnwidth,angle=0]{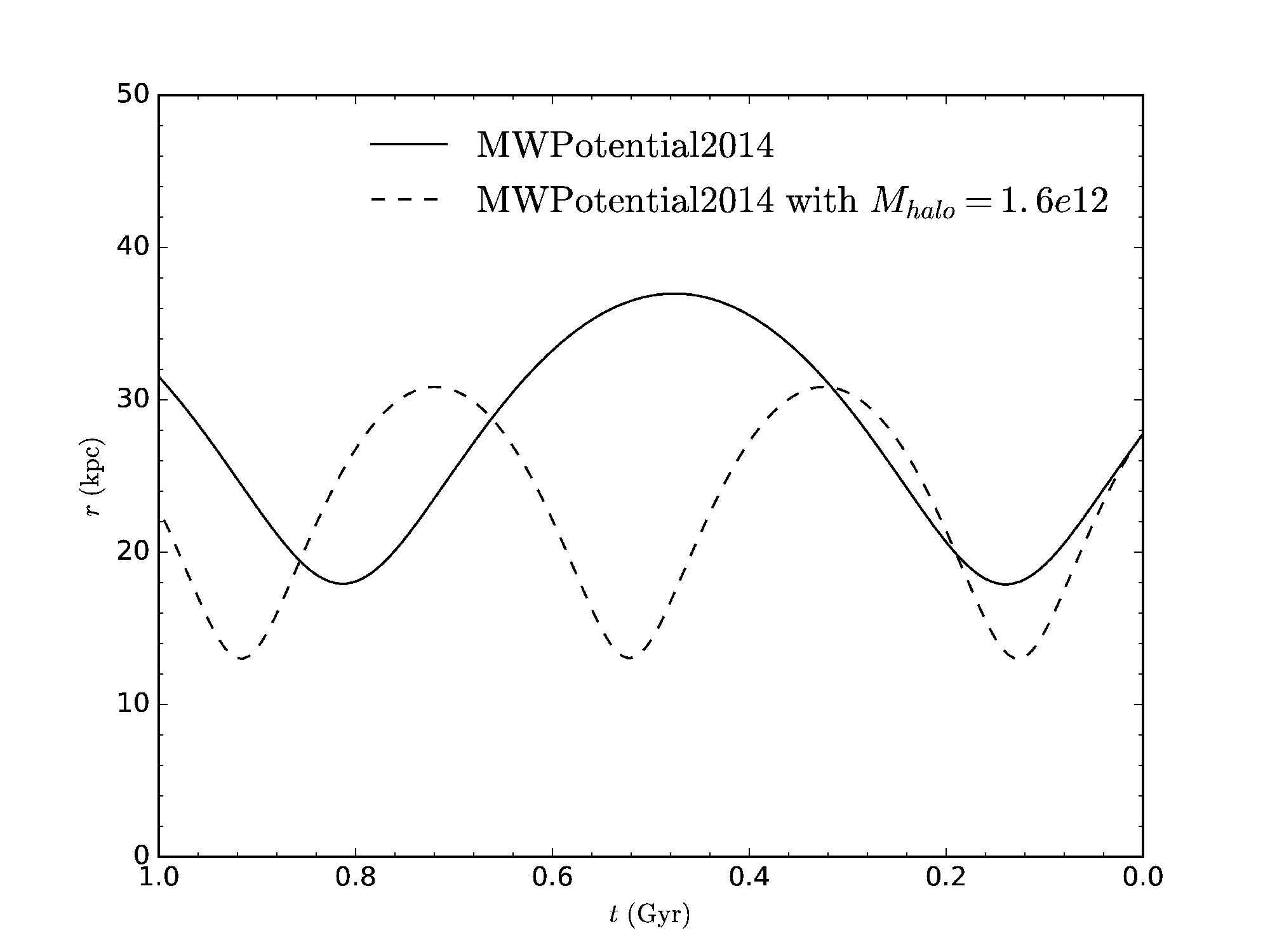}
 \caption{Orbit of Segue~1 for the last  Gigayear of lookback time. The two orbits shown are calculated for the measured proper motion using two different potentials. The line is in the ${\tt MW2014}$ potential; the dashed curve is in a similar potential, but with a twice as massive halo. 
 } 
 \label{fig:peri_orbit}
 \end{center}
 \end{figure}

The ${\tt MW2014}$ potential has a virial halo mass of $0.8\times10^{12}$ M$_\odot$, which is small compared to the many recent measurements; e.g. \citet{Marel_12b} and \citet{Boylan_13} obtain a halo mass of $1.6\times10^{12}$ M$_\odot$. We therefore use as a potential variant a modified ${\tt MW2014}$ potential in which the disk and bulge remain the same but the halo mass is increased by a factor of two, as used in \citet{Fritz_17}. This leads to 13.0$^{+10.6}_{-6.8}$ kpc for the pericenter and 30.9$^{+5.2}_{-3.2}$ kpc for the apocenter. Tracks for the two potentials are shown in Figure~\ref{fig:peri_orbit}. The true halo potential is likely in between these two options; e.g., the review of \citet{Bland_16} obtain a virial mass of $1.3\pm0.3\,10^{12}$M$_\odot$. 
We do not explore different concentrations because the two cases, which both use $c=15.3$, explore the possible range of observed $V_\mathrm{rot}$ in the solar neighborhood of 220 to 256 km~s$^{-1}$ \citep{Bovy_14,Bland_16}. Models with smaller concentration and smaller halo mass lead to values of $V_\mathrm{rot}$ that are too small. Models with smaller concentration and larger halo mass are possible, but lead to apo/peri centers which lie between the two cases. Therefore we use the average of the two discussed cases, which represent the extreme ends of possible values, in the following.

That implies 15.4$^{+10.1}_{-9.0}$ kpc for the pericenter and 33.9$^{+21.7}_{-7.4}$ kpc for the apocenter. The uncertainty also includes the scatter between the two cases but is dominated by the uncertainty of each case.

The most likely orbital pole of Segue~1 is at  $l=23^\circ$ and $b=50^\circ$. 
Due to the large proper motion uncertainties, other uncertainty sources are negligible. Thus its uncertainty bar follows a great circle as shown in Figure~\ref{fig:orb_pol}.

  \subsection{Comparison with the literature and other satellites} \label{sec:other}

  We now compare the orbit of Segue~1 with the orbits of other satellites to address the question of whether Segue~1 could have once been a satellite of one of these galaxies. We therefore concentrate on satellites with a tangential motion measurement, since otherwise the orbit is too uncertain for a meaningful comparison. Besides Segue~1, there are currently 11 galaxies with a motion measurement in the Milky Way halo; see \citet{Pawlowski_13} for an overview. That list consists of the Magellanic Clouds, Sgr Dwarf, and all classical dwarfs (Fornax, Carina, Sextans, Sculptor, Leo I, Leo II, and as the faintest two, Draco and Ursa Minor, with M$_V=-8.8$). 
  Many of the satellites of the Milky Way cluster in one plane \citep{Lynden-Bell_76,Kroupa_05,Pawlowski_15}, known as the vast-plane-of-satellites (VPOS). How significant this plane is,  
 is still up for debate, but until now most satellites with proper motions are consistent with being members \citep{Pawlowski_13,Piatek_16,Sohn_17}. The orbital pole of the VPOS is also close to the orbital pole of many satellites.

\begin{figure*}
 \begin{center}
   \includegraphics[width=2.14 \columnwidth,angle=0]{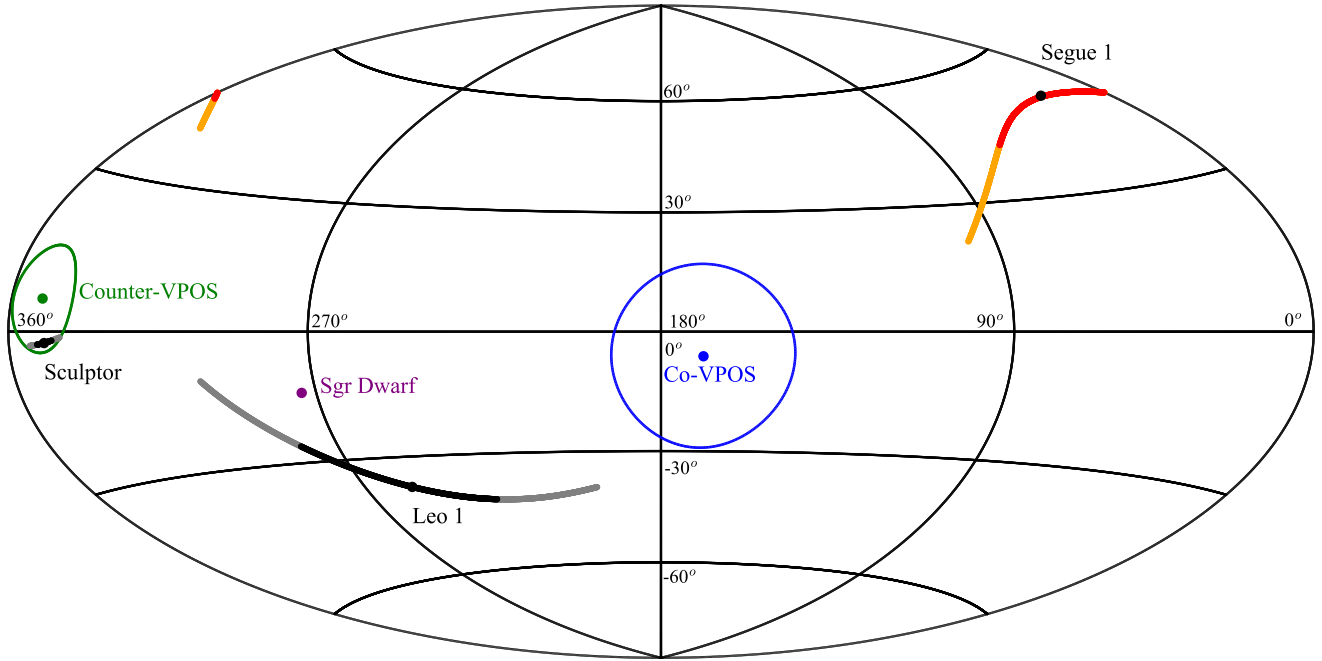}  
 \caption{Orbital pole of Segue~1 and other galaxies. The dark band shows the 1$\,\sigma$ interval of the pole, the lighter the 2$\,\sigma$ interval of the pole  We also show the orbital pole of the VPOS \citep{Pawlowski_15} co-rotating with the Magellanic Clouds (Co-VPOS) and counter-rotating with the Magellanic clouds (Counter-VPOS). We also show the galaxies which are inconsistent with the co-rotating VPOS: Sculptor's pole \citep{Sohn_17} is consistent with the counter-rotating case. The galaxies Sgr Dwarf \citep{Law_10} and Leo I \citep{Sohn_13} probably have orbital poles distinct from either VPOS.
 } 
 \label{fig:orb_pol}
 \end{center}
 \end{figure*}

 For the pole we adopt VPOS$+$new$-$4 measured from \citet{Pawlowski_15}, which includes all of the known galaxies at that time minus 4 outliers. It results in a pole at $(\ell,b) = (169.4^\circ, -6.1^\circ)$, and Figure~\ref{fig:orb_pol} shows clearly that the orbital pole of Segue~1 is different from the VPOS poles. That is true for both options: for galaxies which orbit the Milky Way in the same sense as the LMC, and for galaxies with the opposite sense of rotation, like Sculptor \citep{Sohn_17}. To quantify the mismatch we use Monte Carlo simulations in which we add Gaussian uncertainties to our measurements. We obtain that only in 0.1\% of all cases the pole of Segue~1 is closer than $40^\circ$ to the co-rotating pole of the VPOS; in only 0.3\% of all cases it is closer than $40^\circ$ to the counter-rotating pole of the VPOS. The numbers are not sensitive to which precise VPOS pole is used. That can also be seen by the fact that our proper motion of  $\mu_\alpha=-0.37$ mas/yr and $\mu_\delta=-3.39$ mas/yr is outside of the prediction by \citet{Pawlowski_13}, which uses an older sample to define the VPOS. They predicted for Segue~1 in the co-rotating  case $\mu_\alpha=-0.36$ to 2.37 mas/yr and $\mu_\delta=-1.43$ to 2.49 mas/yr, and in the counter-rotating case $\mu_\alpha=-1.03$ to $-3.76$ mas/yr and $\mu_\delta=-2.39$ to $-6.31$. 
   
   Several  satellites are not in full agreement with the VPOS pole. In the case of Leo I \citep{Sohn_13}, the consistency with the VPOS is borderline, its pole clearly different from the pole of Segue~1 (see Figure~\ref{fig:orb_pol}).  In the case of Ursa Minor I, only one of two measurements \citep{Schweitzer_97,Piatek_05} is roughly consistent with the co-rotating VPOS. However, both are in inconsistent with the pole of Segue~1. Similarly, Sextans' membership in the VPOS is possible with the measurement of \citet{Walker_08}, but the more precise motion of \citet{Casetti_17} makes membership very unlikely. However, both motions result in a pole inconsistent with the pole of Segue~1.
A satellite clearly outside of the VPOS is Sgr Dwarf, which has an orbital pole at $l/b=274/-14$ \citep{Law_10} (That pole is obtained by modeling its stream; most proper motions, see the compilation in \citet{Pawlowski_13} and \citet{Massari_13}, give consistent poles.).
 This is clearly inconsistent with Segue~1. Thus, the poles of all 11 dwarf galaxies with proper motions are likely inconsistent with the pole of Segue~1.

  The presence of the Magellanic Clouds complicates the picture by exerting additional forces \citep{Gomez_15,Sohn_17}. We estimate the force ratio of the LMC relative to the force of the MW by backtracking the LMC and Segue~1 for 1 Gigayear. For the LMC we use the motion of \citet{Kallivayalil_13}. 
  When constant rotation curves are assumed for the LMC and MW, the force ratio depends only on the distance of Segue~1 to the centers of both galaxies. For the MW we adopt a rotation velocity of $\sim$200 km~s$^{-1}$ within the smallest approach of Segue~1 \citep{Kuepper_15}, and for the LMC we adopt a value of 90 km~s$^{-1}$ \citep{Marel_14}. Thus the force by the LMC is at most 5\% of the force of the MW and we can ignore the LMC for Segue~1 in contrast to Sculptor, Draco, and other more distant galaxies. Thus, the orbital pole of Segue~1 is rather robust. Furthermore, most of the 11 dwarf galaxies are at larger distances from the Milky Way than Segue~1, which also makes it less likely that Segue~1 was once a satellite of these galaxies.
  
 On the other hand, the Sgr Dwarf's pericenter is about 20 kpc, similar to that of Segue~1. However, because the difference in the orbital poles, in particular, is large, a common origin is probably excluded. 
 In conclusion, Segue~1 is probably unassociated with any classical dwarf or Magellanic galaxy.

  \citet{Dominguez_16} predicted a proper motion for Segue~1 under the assumption that it is a currently tidally disrupted star cluster which produces the East-West stream detected by \citet{Niederste_09,Bernard_16}. Under these assumptions they predict a proper motion of $\mu_\alpha=-0.19$ mas/yr and $\mu_\delta=-1.9$ mas/yr. Since this motion, which corresponds to a pericenter of 3 kpc and an apocenter of 32 kpc, is not consistent with our motion, some part of the model is ruled out. We used our ${\tt galpy}$ modeling to check whether the orbital path of Segue~1 is along that stream. We obtain that the observed stream is clearly inconsistent with the orbit of Segue~1, which is nearly exactly North-South. That rules out an association of Segue~1 with the stream. \citet{Simon_11} derive that a pericenter of less than about 4 kpc is necessary for tidal disruption.
  Thus, the fact that our proper motion requires a  pericenter larger than 6 kpc makes it less likely that Segue~1 is a nearly disrupted star cluster. It is very likely a galaxy, as was already suggested by its Fe spread \citep{Simon_11,Frebel_14}. 
  
  Given its orbit there are two possibilities for the origin of the dwarf galaxy Segue~1. Firstly, it could have formed with approximately its current total mass, alone, and stayed at about the same distance from the Galactic Center as today since low mass objects are not responsive to dynamical friction \citep{Chandrasekhar_43,Boylan_08}. That case requires that it was never destroyed by the Milky Way, despite always having been rather close. In the second case, Segue~1 was accreted as a satellite of a more massive dwarf galaxy, which due to its larger mass was more sensitive to dynamical friction. This option reduces the time for how long Segue~1 was close to the Milky Way. 
  However, since there is no massive satellite with a matching orbital pole, the infall would have had to have happened some time ago.
To look closer into these two cases, we now compare with cosmological simulations.

\subsection{Insight from cosmological simulations} \label{sec:simulation}
To infer more about the history of Segue~1, we now use the ELVIS simulations of  \citet{Garrison_14}. They are a set of high resolution dark matter only (DMO) simulations of 48 Milky Way sized halos. 
 To draw Segue~1 analogs from the simulation we use the properties of Segue~1: halo mass, distance, radial and tangential velocities. We include all satellites within 2 $\sigma$ of the preferred values and we weight subhalos by their difference from the measured value for Segue 1, assuming Gaussian probabilities in all 4 quantities.
For the halo mass of Segue~1 we use 10$^8$ M$_\odot$, motivated by the abundance matching relation from \citet{Garrison_14}, assuming 0.5 dex scatter for $\sigma$.
This results in a sample of 29 Segue~1 analogs. Of these, 1 was once a satellite of a currently surviving dwarf galaxy before infall into the Milky Way. Since these galaxies are usually rather massive (LMC like, see \citep{Wetzel_15b}), they are excluded by the orbital pole difference between Segue~1 and other dwarf galaxies (Section~\ref{sec:other}). We therefore exclude this case and have 28 remaining. 
To obtain cumulative fractions and weights of halo properties, we adjust the probabilities 
such that the sum of all weights is 1. 
(See Wetzel \& Tollerud, in prep, for more details on this methodology.)

Because the ELVIS simulations incorporate only dark matter, they do not incorporate the effects of the central galaxy disk, which affects the population of surviving subhalos \citep{Garrison_17}. Simulations which contain baryons are now possible, e.g. Latte \citep{Garrison_17} and APOSTLE \citep{Sawala_16}, 
but they are too expensive to run at a number sufficient to sample subhalo orbital histories. 
In \citet{Garrison_17} it is explained that whether a subhalo is affected by baryons depends primarily on a single parameter, the smallest distance ($r_\mathrm{min}$) experienced by a halo. 
Thus, we can model the effects of baryons by using this information. 
To establish how likely a subhalo is to be destroyed by the central galaxy we obtain the number of surviving subhalos in bins in log($r_\mathrm{min}$) for two simulations using the same starting conditions with and without baryons added. This uses the data presented in Figure~5 of \citet{Garrison_17}. As in that work we use two simulations, m12f and m12i, to establish uncertainties and to increase the sample. We then fit the binned data by a hyperbolic tangent. For the fit we ignore the data beyond 100 kpc, since that is well outside of the $r_\mathrm{min}$ of Segue~1 analogs. The fit and the $r_\mathrm{min}$ of the Segue~1 analogs is then used to derive the survival probability of each analog. The analogs to Segue 1 in ELVIS are destroyed with 65\% probability, implying that Segue~1-like galaxies are often but not always destroyed. Thus, given that observations are biased towards close galaxies, it is not surprising that a Segue~1-like galaxy has been discovered. The survival probabilities are then multiplied by the DMO weights. To bring the total weights back to 1 we then divide through by the sum of all weights.  After the adjustment, the range in weight is between 0.06\% and 12.6\%. 

  \begin{figure}
 \begin{center}
     \includegraphics[width=0.70 \columnwidth,angle=-90]{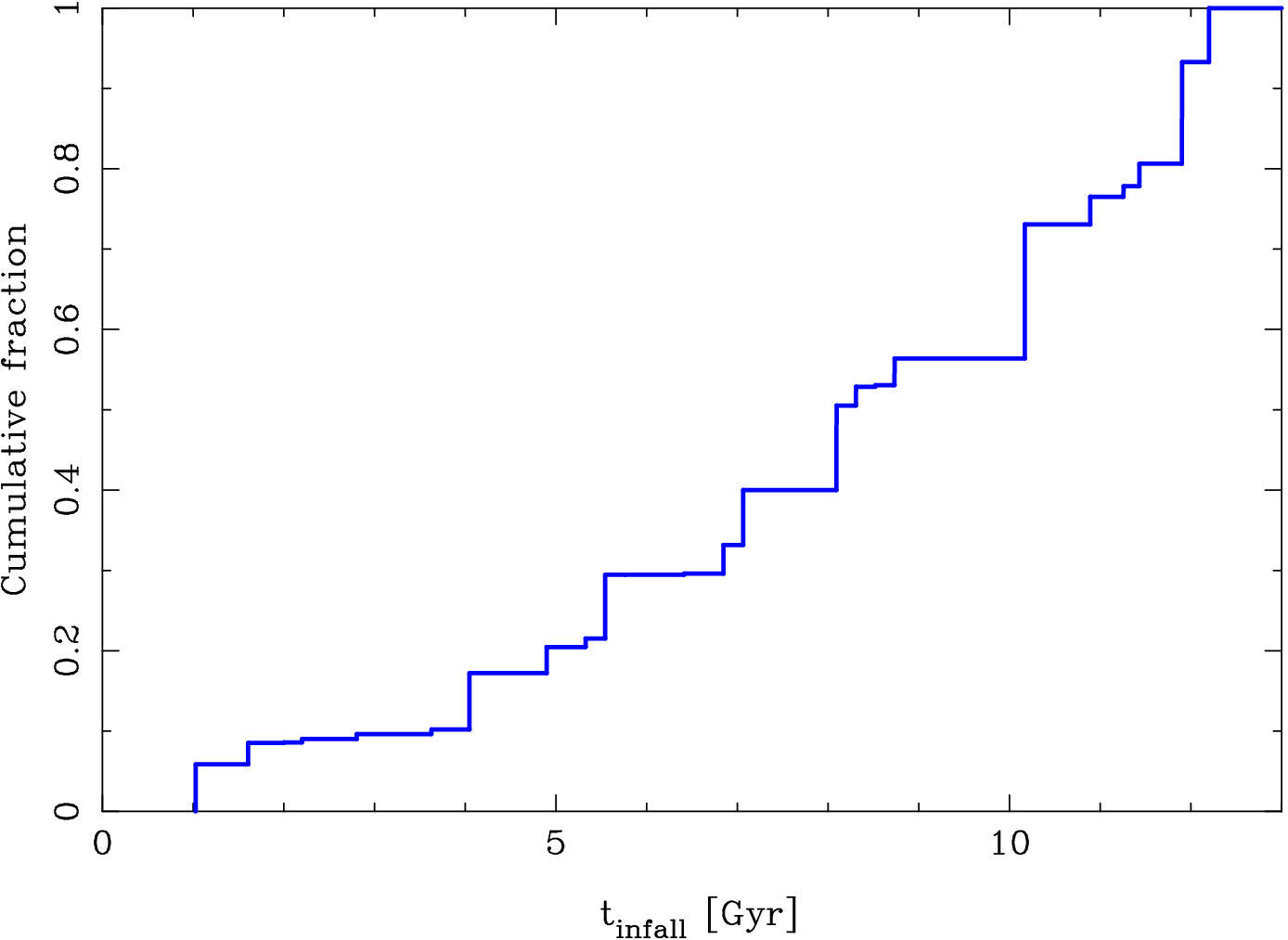}
     \includegraphics[width=0.70 \columnwidth,angle=-90]{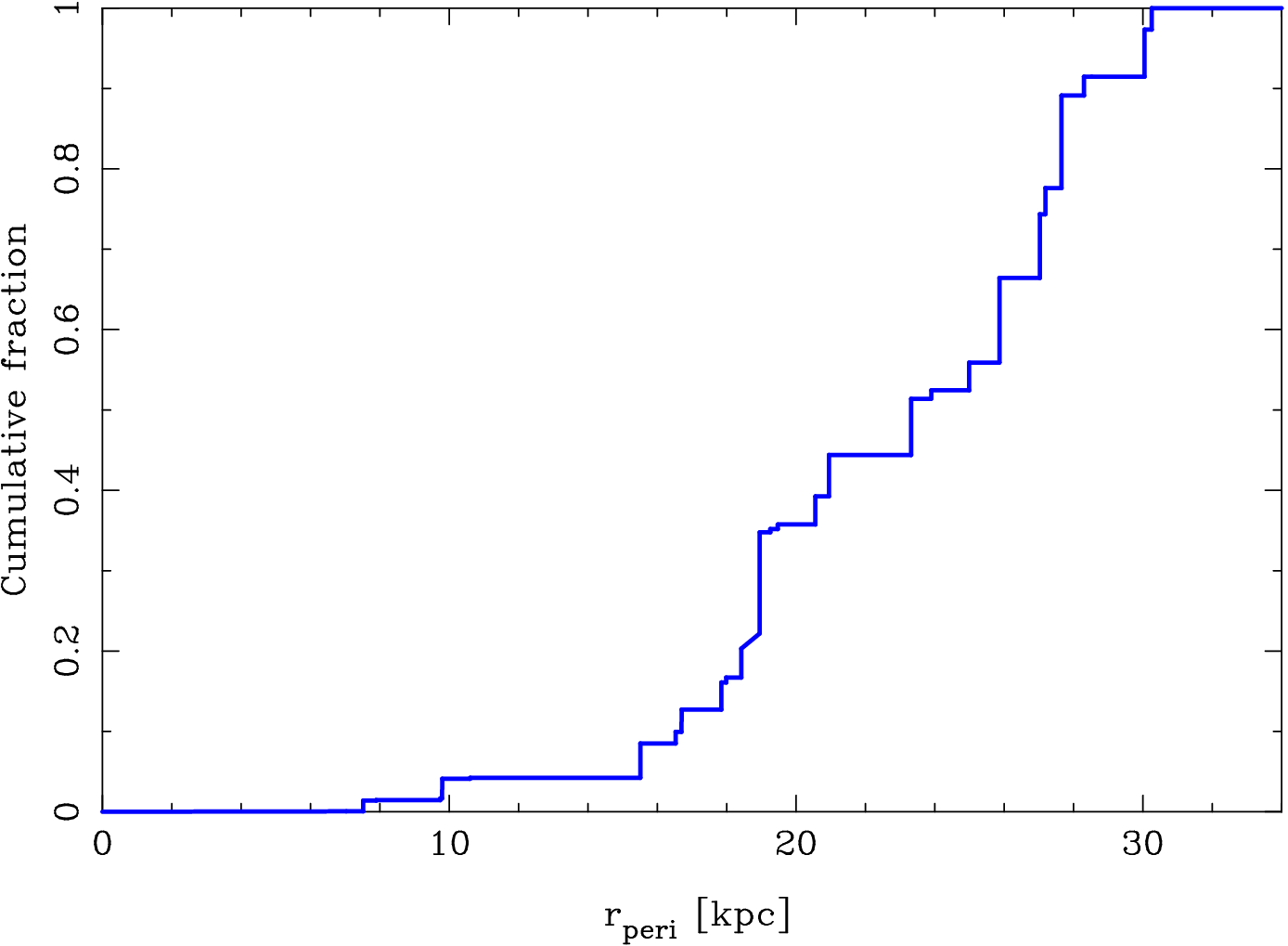}
    \includegraphics[width=0.70 \columnwidth,angle=-90]{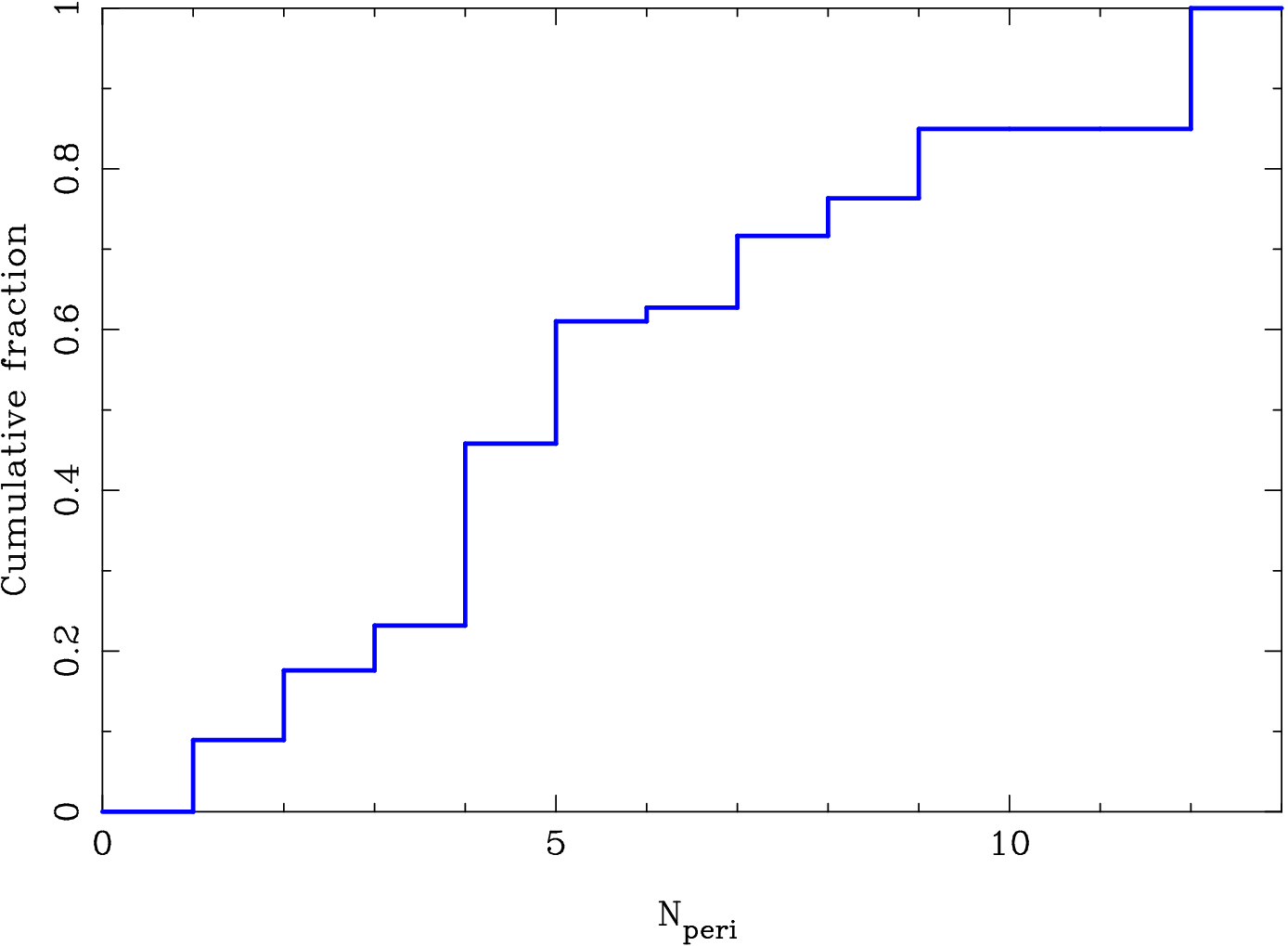}
 \caption{Properties of Segue~1 analogs in cosmological simulations. We use the ELVIS simulations \citep{Garrison_14}, weighting
the subhalos by their probability of surviving tidal disruption by the central galaxy disk, based on the results of \citet{Garrison_17}. Thus we effectively account for the effects of the central galaxy disk.
 Top: infall time of the Segue~1 analog into the Milky Way analog. Middle: smallest distance relative the Milky Way (physical) of the Segue~1 analog. Bottom: number of pericenters of the Segue~1 analog relative to the Milky Way.
 }
 \label{fig:sim_plots}
 \end{center}
 \end{figure}

In Figure~\ref{fig:sim_plots} we show cumulative distributions for the time since first infall of the Segue~1 analogs into the Milky Way, the 
pericenter relative to the center of the Milky Way, and the number of pericenters experienced since first infall into the Galaxy. The median values and 1 sigma ranges are 8.1$^{+3.6}_{-4.3}$ Gyrs, 22.8$^{+4.7}_{-4.8}$ kpc and 4$^{+5}_{-2}$ orbits. The smallest distance to the Milky Way is increased compared to the DMO case, where the median value is 18.9 kpc, because halos passing close to the Milky Way are likely to be destroyed by the Galactic disk. Because in the DMO case the innermost halos are also more likely to be destroyed, analogs in both simulations have preferentially larger pericenters than the unweighted observations (Section~\ref{sec:orbit}). 
Additionally, the infall time is preferentially early. Analogs which where already a satellite of another galaxy when accreting have a median $t_\mathrm{infall}=12$ Gyrs, while the others have $t_\mathrm{infall}=8$ Gyrs. 
Spectroscopic properties favor very early formation of Segue~1 possibly at $z\sim10$ \citep{Webster_16}. 
Since this is before all infall times derived in this work, the spectroscopy does not provide additional constraints
on the infall history of Segue~1.

In general, about one third of all satellites of MW-like galaxies were once a satellite of another galaxy before accretion onto the Galaxy \citep{Wetzel_15b}. Since most of these host satellites have a halo mass near 10$^{11}$ M$_\odot$, i.e. near that of the LMC (M$_V=-18.1$), the fact that we can only exclude galaxies with  M$_V<-8.8$ (Section~\ref{sec:other}) still means that we can exclude more than 90\% of the cases in which the former host survived. Thus, we assume that we can exclude all cases in which a former host galaxy survived until today.  
When we use all Segue 1 analogs with no weighting by the probability
of tidal disruption by the central baryonic disk, we obtain that 37\% of all analogs were once a satellite of another Galaxy, which is now destroyed, before they were accreted to the MW. This fraction is higher than the general $1/3$ \citep{Wetzel_15b}, we speculate because tightly bound satellites like Segue~1 are more likely delivered by massive hosts, since they will experience stronger dynamical friction and get close to the Galaxy.
The fraction of former satellites whose host did not survive until today is reduced to 25\% when we include the effects of the central Galaxy disk and exclude the cases in which the host survived until today.  Also, in principle an already destroyed host can leave a visible imprint on the Milky Way as a sign of a past merger \citep{Quinn_93,Kauffmann_96}. However, in all our cases the host infall happened rather early, the  most recent case is 7.1 Gyrs ago. Thus, the fact that the Milky Way is rather quiet \citep{Freeman_02} and that the visible disturbances can probably be explained by known galaxies \citep{Laporte_17} do not give relevant additional constraints. Therefore both cases, accretion alone and accretion as a satellite of a now destroyed galaxy are possible, but the first option is preferred.

\section{Summary} \label{sec:summary}

\begin{enumerate}
\item $\,$We use the SDSS catalog and dedicated LBC/LBT images to measure the proper motion of the ultra-faint dwarf spheroidal Segue~1.
\item $\,$We measure the motion of Segue~1 members relative to faint background galaxies. These galaxies are of very high astrometric quality because they are required to be morphologically consistent in 4 different images and three different bands. 
\item $\,$For our proper motion we consider, in particular, the following effects: distortion correction, differential chromatic refraction, Segue~1 membership uncertainty and random uncertainty of these and the reference galaxies.
\item $\,$We obtain the first proper motion of Segue~1: $\mu_{\alpha}\,\cos(\delta) = -0.37\pm0.57$ mas yr$^{-1}$ and $\mu_{\delta} =-3.39\pm0.58$ mas yr$^{-1}$. 
 Combining this with the known line-of-sight velocity produces a Galactocentric V$_\mathrm{rad}=84\pm9$ and V$_\mathrm{tan}=164^{+66}_{-44}$ km~s$^{-1}$.
\item $\,$Considering the uncertainties of the Segue~1 phase space properties (in particular proper motion and distance) and the uncertainties in the potential of the Milky Way we obtain 15.4$^{+10.1}_{-9.0}$ kpc for the pericenter and 33.9$^{+21.7}_{-7.4}$ kpc for the apocenter.
\item $\,$The fact that the pericenter and apocenter are not very close to the Milky Way further strengthens the case for Segue~1 being a galaxy instead of a star cluster in the process of disruption, since tidal disruption is too weak at the distances of Segue~1.
\item $\,$The most likely orbital pole of Segue~1 is at  $l=24^\circ$ and $b=50^\circ$. While there is some uncertainty, it is clearly different from the orbital pole of all galaxies with tangential velocities. That includes all classical dwarfs and the more massive satellites. Thus, it is very unlikely that Segue~1 was once a satellite of a more massive known satellite.  
\item $\,$We use cosmological 
simulations to get properties of Segue~1 analogs.  We obtain that Segue~1 probably accreted long ago, 8.1$^{+3.6}_{-4.3}$ Gyrs, and experienced about 4 pericentric passages since then. The pericenter of the analogs is 22.8$^{+4.7}_{-4.8}$ kpc. This is mostly larger than the pericenter of the orbit, because galaxies with a small pericenter are more likely to be destroyed by the Galaxy, especially its disk. There is a 75\% probability that Segue~1 accreted alone and  a 25\% probability that it was once the satellite of a massive LMC-like galaxy, now destroyed. In the latter case the median accretion time of both galaxies to the Milky Way was 12 Gyrs ago.

\end{enumerate}

\acknowledgements
This work was supported by the NSF CAREER award 1455260.\\
This research was supported in part by the National Science Foundation under Grant No. NSF PHY-1125915, and by the hospitality of the Kavli Institute for Theoretical Physics at the University of California, Santa Barbara.\\
AW was supported by NASA through grants HST-GO-14734 and HST-AR-15057 from STScI.\\
We thank Shea Garrison-Kimmel for making the ELVIS simulations publicly available and for sharing data from Fig.~5 of \citet{Garrison_17}.\\
We thank Jo Bovy for advice regarding ${\tt galpy}$.

  \facility{SDSS, LBT(LBC)} 
  \software{ dpuser, SExtractor \citep{Bertin_96},  PSFeX \citep{Bertin_11}, \texttt{galpy} \citep{Bovy_14b}, mpfit \citep{Markwardt_09}, ELVIS \citep{Garrison_14}}

\bibliography{mspap}

\end{document}